# Hiding a Realistic Object Using a Broadband Terahertz Invisibility Cloak

Fan Zhou[1]*, Yongjun Bao[1]*, Wei Cao[2]*, Colin T. Stuart[1], Jianqiang Gu[2], Weili Zhang[2], and Cheng Sun[1]

[1] Mechanical Engineering Department, Northwestern University, Evanston, Illinois 60208, USA.
[2] School of Electrical and Computer Engineering, Oklahoma State University, Stillwater, Oklahoma 74078, USA.

* These authors contributed equally to this work

The invisibility cloak has been a long-standing dream for many researchers over the decades. The introduction of transformational optics has revitalized this field by providing a general method to design material distributions to hide the subject from detection[1,2]. By transforming space and light propagation, a three-dimensional (3D) object is perceived as having a reduced number of dimensions, in the form of points, lines, and thin sheets, making it "undetectable" judging from the scattered field[2-4]. Although a variety of cloaking devices have been reported at microwave and optical frequencies[5-10], the spectroscopically important Terahertz (THz) domain[11-13] remains unexplored. Moreover, due to the difficulties in fabricating cloaking devices that are optically large in all three dimensions, hiding realistic 3D objects has yet to be demonstrated. Here, we report the first experimental demonstration of a 3D THz cloaking device fabricated using a scalable Projection Microstereolithography process. The cloak operates at a broad frequency range between 0.3 and 0.6 THz, and is placed over an α-lactose monohydrate absorber with rectangular shape. Characterized using angular-resolved reflection THz time-domain spectroscopy (THz-TDS), the results indicate that the THz invisibility cloak has successfully concealed both the geometrical and spectroscopic signatures of the absorber, making it undetectable to the observer.

The visual appearance of objects is determined by the extent to which they modify light due to geometric scattering and material absorption. Thus, in order to hide the object from being detected, the invisibility cloak must conceal the changes to the light in both geometry and spectrum. The introduction of transformational optics provides a generalized method to design the distributions of material constitutive parameters that utilizes the form invariance of Maxwell's equations[1,2]. However, the cylindrical cloak relies on strong materials dispersion to provide the necessary dielectric singularity without violating physical laws. Thus, the cloaking devices all exhibit high loss and could only be effective at a single wavelength, making them incapable of cloaking in the spectrum domain[1,2]. To overcome this constraint, the quasiconformal mapping (QCM) technique was used to eliminate the need for the dielectric singularity[4,5,14] and significantly mitigated the anisotropy of the material, allowing the realization of a broadband invisibility cloak using only isotropic dielectrics. Particularly, a carpet cloak using QCM was proposed to transform one curved reflective surface into a flat sheet, which required only a moderate variation of the permittivity and permeability. Manually assembling a series of non-resonant elements would be the simplest way to build the carpet cloak at microwave frequencies[7,10]. However, such a manual assembly strategy becomes inapplicable at high frequencies due to the reduction of the element size. Carpet cloak devices at infrared frequencies[6,8] were instead fabricated using conventional top-down nanofabrication techniques, such as e-beam lithography and focused ion beam milling, which critically constrain the cloak structure in the third dimension. Thus, these cloaking devices were only realized in the 2D waveguiding modes. Recently, Ergin *et al.* has successfully demonstrated the first 3D cloak by direct writing a 10 μm thick wood pile structure inside photoresist (PR) layer using a commercial system[9]. Limited light penetration into the absorbing PR layer makes it difficult to fabricate thicker samples. On the other hand, despite the exciting new discoveries in microwave and optical frequencies, it should be noted that the spectroscopically important THz gap has yet to be explored. The specific vibrational, rotational, and translational responses of materials within THz region carry unique molecular signatures that are generally absent in optical, microwave, and X-ray frequencies[11-13]. The transformational optics will provide unprecedented capabilities in manipulating the THz wave propagation for spectroscopic analysis. Cloaking of realistic objects under THz is a representative example and is inspiring rapidly increasing interest. However, the lack of scalable microfabrication technologies for the creation of an optically large 3D cloaking device containing deep sub-wavelength features remains as the major obstacle towards the realization of a 3D THz cloak.

Here, we report the first experimental demonstration of a broadband THz cloaking device that can be placed over a realistic 3D object to conceal its geometrical and spectroscopic features simultaneously. The refractive index distribution in the *x-y* plane is obtained using QCM for transverse magnetic (TM) waves[4]. The Modified-Liao functional is applied to control each parallelogram in the mesh-generation process to reduce the material's anisotropy. In this work, we design a triangle-shaped carpet cloak with a width of 8.52 mm and height of 4.26 mm. The bump located at the bottom surface is defined by a cosine function with a height of 0.213 mm and a width of 4.26 mm. After conformal mapping, the permittivity of the ground-plane cloak is found in the range of 1.02 to 2.86 with a corresponding background of permittivity $\varepsilon_{ref}$=1.47 ($n_{ref}$=1.21). The calculated anisotropic parameter is 1.04, thus the cloak pattern can be treated as an isotropic medium. Under effective media approximation, the permittivity distribution can be realized by creating sub-wavelength holes of variable diameters inside the polymer host. For transverse magnetic (TM) waves, the filling ratio of air holes $f$ can be calculated using the effective media theory:

$$f = \frac{\varepsilon_d/\varepsilon_{eff} - 1}{\varepsilon_d/\varepsilon_{air} - 1}$$

where $\varepsilon_{eff}$ is the effective permittivity of the cloak pattern, $\varepsilon_{air}$ and $\varepsilon_d$ are the permittivity of air and the host polymer materials, respectively. Using a unit cell size of 85.2 μm, the corresponding hole side length ranges from 10 μm to 71 μm. The design is extruded along the $z$ axis to form the 3D cloaking structure (Fig. 1a). The characteristic of such a cloak design consisting of microscopic hole structures is evaluated using numerical simulation (COMSOL Multiphysics) for a bumped surface without and with a cloak, as shown in Figs. 1b and 1c, respectively. The simulations show the magnetic field component in the z direction at 0.6 THz. The THz wave reflected from the uncloaked bump exhibits two distinct peaks and one minor peak in between. In contrast, the reflection from the cloaked bump shows the expected flat wavefront, as if the bump does not exist. Thus, the object placed underneath the reflective bump can be cloaked.

We employed a highly scalable and parallel 3D microfabrication technique, named projection microstereolithography (PμSL), for fabrication of the 3D cloaking device[15]. In contrast to the direct writing process fabricating the 3D structure in a point-by-point scanning fashion, PμSL uses a high resolution dynamic mask containing 1,470,000 (1,400×1,050) pixels, to fabricate a whole 2D pattern in a single exposure (Fig. 2a). The desired permittivity profile is obtained by using sub-wavelength unit cells containing through holes with varying dimension under effective media approximation.

In this work, the unit cell has a width of 85.2 μm (~12 pixels) in the *x-y* plane and 20 μm in the *z* direction, which are much smaller than the vacuum wavelength of the THz wave to the order of 500 μm (Fig. 2a). Although the pixels on the dynamic mask are presented in the discrete form, tuning the intensity of individual pixels at 256 grayscale levels allows the fabrication of holes with sub-pixel scale dimensional accuracy (Supplementary Information, Fig. S1). It is then possible to create the desired refractive index distribution with smoother transitions in space to minimize the scattering of the THz wave. The designed mask pattern in a digitized form is used for constructing the 3D cloaking structure in a layer-by-layer fashion. In analog to the natural crystals formed by the 3D assembly of atoms, the cloaking structure is constructed from bottom up to form the assembly of unit cells in 3D with varying effective permittivity. As shown in Fig. 2b, the fabricated triangular cloaking structure has a total thickness of 4.4 mm, comprised of 220 layers of 20 μm thickness. The distribution of the varying hole geometry can be clearly identified in the SEM images shown in Fig. 2b. Finally, a 200 nm-thick gold layer is deposited on the bottom to form a reflective surface. The space underneath the bump is designated as the cloaked region.

Angular resolved reflection THz-TDS (Supplementary Information, Fig. S3) was employed to assess the performance of the cloaking sample[16]. The photoconductive switch-based THz-TDS system was optically gated by 30 fs, 800 nm optical pulses generated from a self-mode-locked Ti:sapphire laser. The THz radiation emitted from a GaAs transmitter was spatially gathered by a hyper-hemispherical silicon lens and then collimated into a parallel beam before entering the cloaking sample at a 45° incident angle. The reflected THz signal through the sample was then detected by a mobile silicon-on-sapphire (SOS) receiver optically excited with fiber-coupled femtosecond pulses[17]. The detector was placed 50 mm away from the output interface of the sample and scanned horizontally with a scan step of 0.5 mm to measure the spatial distribution of the THz wave front. The frequency-dependent THz amplitudes at each spatial position are retrieved by Fourier transform of the measured time-domain signals.

In this study, the test sample is made of α-lactose monohydrate, which exhibits a resonant attenuation signature at 0.53 THz due to the presence of collective vibrational transition modes[18]. The α-lactose monohydrate powder (Sigma-Aldrich) is pressed into a thin rectangular shape (1 mm × 0.12 mm × 4 mm) on a reflective substrate, such that it can fit under the bump. The measurement of the reflected THz wave is performed over a broad frequency range of 0.3-0.6 THz on a set of four cases: (I) flat reflective surface, (II) exposed lactose on reflective surface, (III) control structure with reflective bump placed on top of the lactose, and (IV) the cloaking structure placed on top of the lactose. As shown in Fig. 3, the horizontal and vertical axes represent angular positions and frequencies, respectively, while the color represents the normalized amplitude of the THz wave. In case (I), the THz wave incident at 45° is reflected from the gold-coated flat surface. The half-space above the reflective bump has an effective refractive index $n_{ref}$. The measured reflected beam is clearly confined in both frequency and space, as the characteristic of a collimated THz beam. In case (II), the reflected beam is split into three peaks due to the geometrical scattering caused by the lactose. In addition, a narrow dip is observed at 0.53 THz, which is coincident with the absorption signature unique to α-lactose monohydrate. To prevent the lactose from being detected, it is possible to hide it under a reflective bump, as shown in case (III). Shielding the lactose from the THz wave causes the spectroscopic signature at 0.53 THz to disappear. However, the reflected THz wave exhibits three peaks, indicating unsuccessful cloaking due to the effect of geometric scattering off the bump. In contrast, the THz cloak shown in case (IV) conceals both the geometric and spectroscopic features of the lactose, which closely resembles the case (I) as a flat reflective surface. In order to better illustrate the differences, the measured THz amplitude at 0.48 THz for the aforementioned four cases is shown in Fig. 4a. The cloak and flat surface plots resemble each other very closely with a single central peak, whereas the graphs for the exposed lactose and reflective bump exhibit two additional side peaks (red and green arrows). Fig. 4b plots the amplitude as a function of frequency at a detector position of -3 degrees. The 0.53 THz absorption dip (red arrow) observed in the exposed lactose case is not present in the cloak case, which again is similar to the flat reflective surface. These results demonstrate that the fabricated THz invisibility cloak has successfully concealed both the geometrical and spectroscopic signatures of the $\alpha$-lactose monohydrate structure.

In summary, this work not only reports the experimental demonstration of hiding a realistic 3D object from THz waves, but more importantly, it offers a technical feasible solution for designing and fabricating 3D THz transformational optics. In contrast to the conventional optical methods, the ability to manipulate THz wave propagation using transformational optics will undoubtedly initiate new possibilities for a variety of intriguing THz applications in quality control, biomedical imaging, disease detection, sensing, and communications..

**METHODS SUMMARY**

**Design and fabrication of the 3D THz Cloak.** The cloak is designed as a 3D solid model. By slicing of the 3D solid model, each 2D slice is displayed on the dynamic mask as a bitmap image. The high-resolution optical image is then projected on the top surface of the photo-curable polymer resin and subsequently solidified, forming a thin polymer layer with the desired shape. Using a computer to synchronize the vertical motion of the sample stage and a slow movie of the sequential 2D slices on the dynamic mask, the 3D cloaking device can be rapidly constructed. The exposure time for each layer is a few seconds, allowing fabrication of the 200-layer 3D structure shown in Fig. 2b of the main text in less than an hour. For the cloak fabrication, the UV curable resin consists of 1,6-Hexanediol diacrylate (HDDA) as the low viscosity monomer and Irgacure 819 as the photoinitiator. Moreover, a critical amount of UV absorber has been mixed with the UV curable resin to control the curing depth. After the fabrication, the sample was rinsed in isopropyl alcohol for 48 hours to remove the remaining polymer inside the holes.

**Supplementary Information** is linked to the online version of the paper at www.nature.com/nature.

**Acknowledgements**

**Author Contributions**: F.Z. conceived and designed the experiments, fabricated the cloaking and control samples; Y.B. calculated the refractive index profile and performed numerical simulations. W.C characterized the performance of cloaking and control samples; All authors discussed and commented on the manuscript.

**Author Information** Reprints and permissions information is available at www.nature.com/reprints. The authors declare no competing financial interests. Readers are welcome to comment on the online version of this article at www.nature.com/nature. Correspondence and requests for materials should be addressed to C.S. (c-sun@northwestern.edu).

**FIGURE LEGENDS**

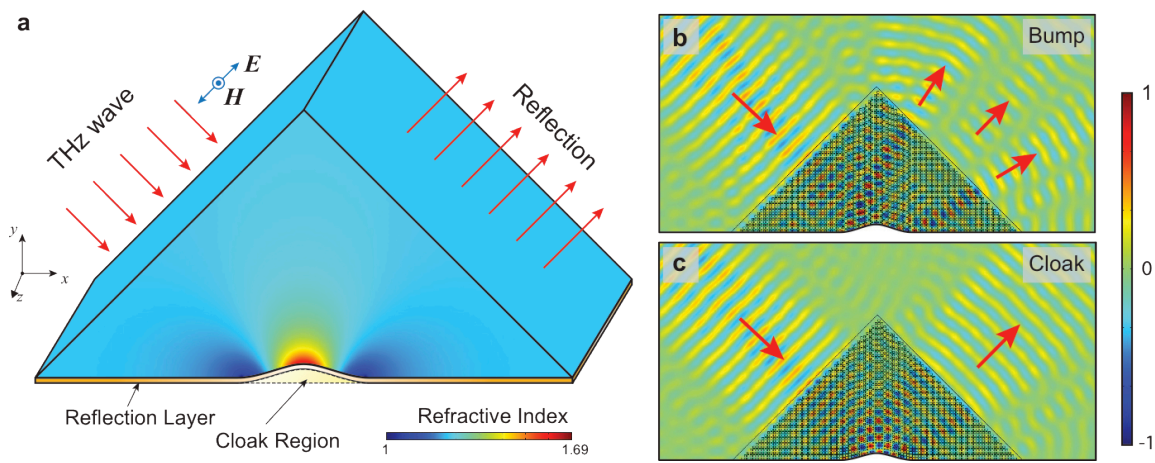

**Figure 1 | Design and simulation of 3D THz cloak. a,** The triangle-shaped cloak structure is obtained by extruding the refractive index profile in the *xy* plane along the *z*-axis. **b,** Numerical simulation using commercial software (COMSOL Multiphysics) confirms the splitting of the wave into three waves for the bump structure with uniform refractive index. **c,** The cloak structure preserves the original shape of the incoming wave. The simulation results show the normalized magnetic field component in the *z* direction at 0.6 THz.

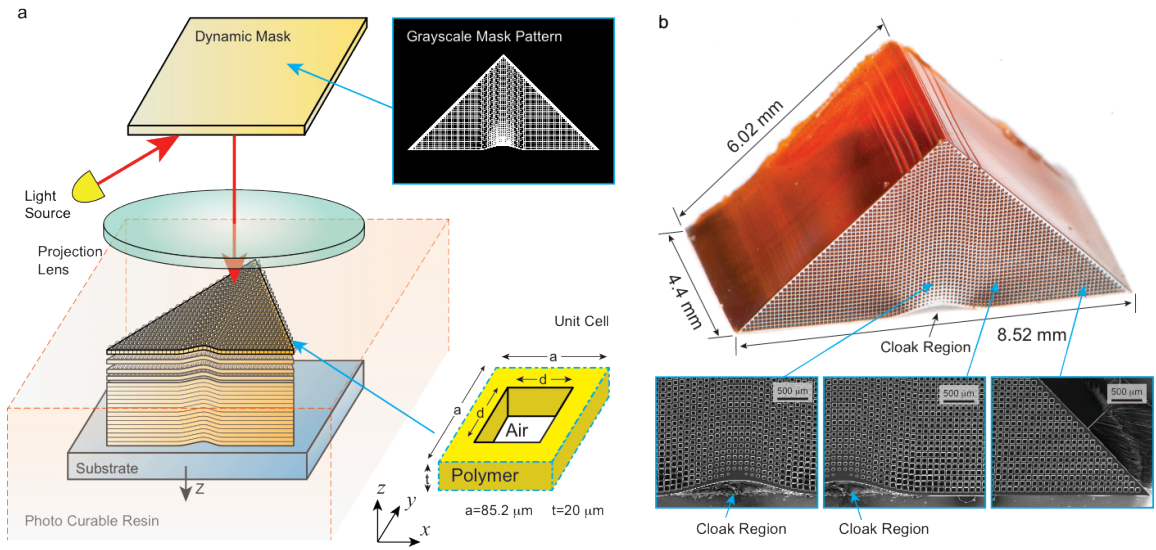

**Figure 2 | Fabrication and characterization of 3D THz cloak. a,** Schematic diagram illustrating projection micro-stereolithography system being used to fabricate 3D cloaking device. The grayscale of individual pixels within each 85.2 x 85.2 μm unit cell can be adjusted so the holes can be fabricated with sub-pixel precision. **b,** Optical and SEM images of fabricated cloaking device. The surface of the device was metalized to enhance the contrast for better representation of the fine features in the images. The gradual change in hole size near the bump can be clearly observed.

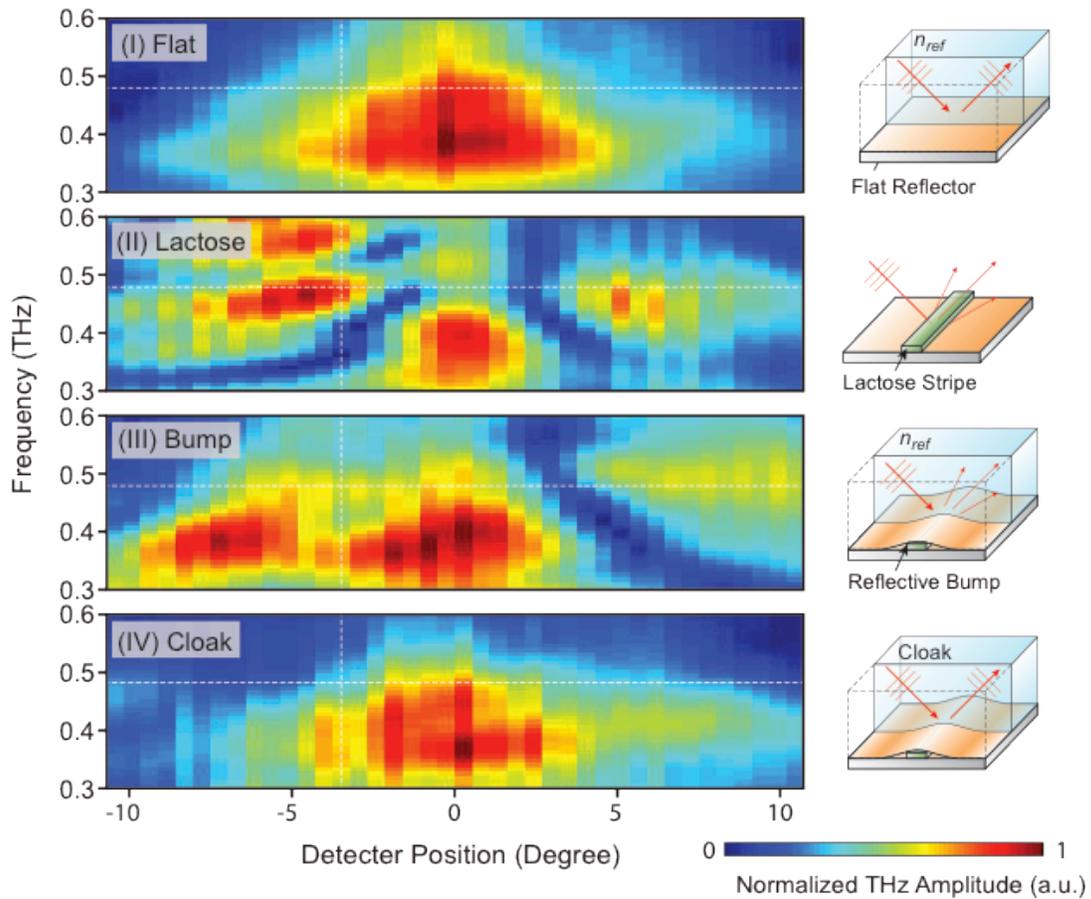

**Figure 3 | Spectra maps of four experimental cases**. (I) Reflective flat, (II) exposed α-lactose monohydrate, (III) reflective bump, and (IV) cloak, are measured using reflection terahertz time-domain spectroscopy. The lactose (II) shows both a scattering effect and absorption. The reflective bump (III) avoids the absorption effect, but is still split into three peaks. The measured spot position of the cloak (IV) and the reflective flat (I) match reasonably well with each other.

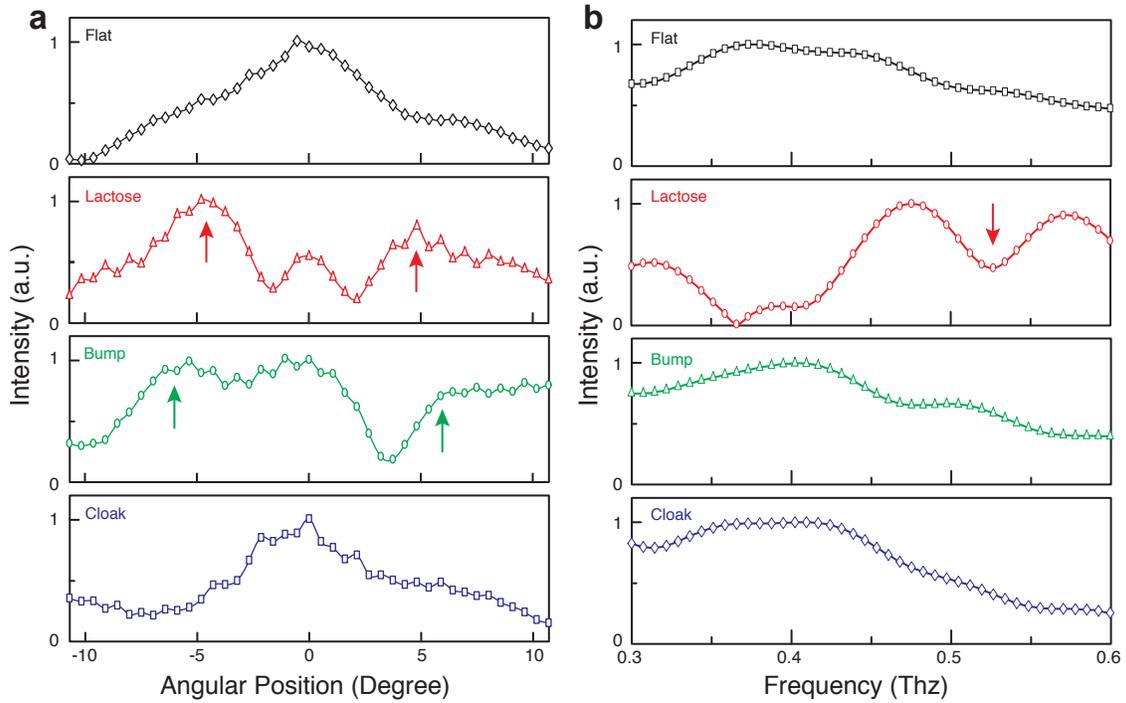

**Figure 4 | Cross-sectional plots of experimental results.  a,** Intensity as function of detector position for four cases at 0.48 THz. The cloak and reflective flat have one similar peak, whereas the lactose and reflective bump exhibit two extra peaks (red and green arrows). **b,** Intensity as function of frequency for four cases at -3 degrees (angle). Obvious absorption occurs at 0.53 THz for the exposed lactose (red arrow).